\documentclass[aps,prd,superscriptaddress]{revtex4-1}
\usepackage{url}
\usepackage{graphicx}
\usepackage{amsmath}
\usepackage[all]{nowidow}
\usepackage[pdftex,bookmarks,hidelinks]{hyperref}
\newcommand{\nue}{\mbox{$\nu_e$}}
\newcommand{\nuebar}{\mbox{$\overline\nu_e$}}
\newcommand{\numu}{\mbox{$\nu_{\mu}$}}
\newcommand{\numubar}{\mbox{$\overline\nu_{\mu}$}}
\newcommand{\nutau}{\mbox{$\nu_{\tau}$}}
\newcommand{\nutaubar}{\mbox{$\overline\nu_{\tau}$}}

\newcommand{\thetatwothree}{\mbox{$\theta_{23}$}}

\newcommand{\dcp}{\mbox{$\delta_{CP}$}}

\newcommand{\sigmahat}{\mbox{$\hat{\sigma}$}}

\makeatletter
\def\and{%                  % \begin{tabular}[t]{c}
  \end{tabular}%
  \hskip 0.01em %
  \begin{tabular}[t]{c}}%   % \end{tabular}
\makeatother
\begin{document}

\title{Experiment Simulation Configurations Approximating DUNE TDR}
%\authorrunning{DUNE Collaboration}
\newcommand{\Amsterdam}{University of Amsterdam, NL-1098 XG Amsterdam, The Netherlands}
\newcommand{\Antananarivo}{University of Antananarivo, Antananarivo 101, Madagascar}
\newcommand{\AntonioNarino}{Universidad Antonio Nari{\~n}o, Bogot{\'a}, Colombia}
\newcommand{\Argonne}{Argonne National Laboratory, Argonne, IL 60439, USA}
\newcommand{\Arizona}{University of Arizona, Tucson, AZ 85721, USA}
\newcommand{\Asuncion}{Universidad Nacional de Asunci{\'o}n, San Lorenzo, Paraguay}
\newcommand{\Athens}{University of Athens, Zografou GR 157 84, Greece}
\newcommand{\Atlantico}{Universidad del Atl{\'a}ntico, Atl{\'a}ntico, Colombia}
\newcommand{\Banaras}{Banaras Hindu University, Varanasi - 221 005, India}
\newcommand{\Basel}{University of Basel, CH-4056 Basel, Switzerland}
\newcommand{\Bern}{University of Bern, CH-3012 Bern, Switzerland}
\newcommand{\Beykent}{Beykent University, Istanbul, Turkey}
\newcommand{\Birmingham}{University of Birmingham, Birmingham B15 2TT, United Kingdom}
\newcommand{\BolognaUniversity}{Universit{\`a} del Bologna, 40127 Bologna, Italy}
\newcommand{\Boston}{Boston University, Boston, MA 02215, USA}
\newcommand{\Bristol}{University of Bristol, Bristol BS8 1TL, United Kingdom}
\newcommand{\Brookhaven}{Brookhaven National Laboratory, Upton, NY 11973, USA}
\newcommand{\Bucharest}{University of Bucharest, Bucharest, Romania}
\newcommand{\CBPF}{Centro Brasileiro de Pesquisas F\'isicas, Rio de Janeiro, RJ 22290-180, Brazil}
\newcommand{\CEASaclay}{CEA/Saclay, IRFU Institut de Recherche sur les Lois Fondamentales de l'Univers, F-91191 Gif-sur-Yvette CEDEX, France}
\newcommand{\CERN}{CERN, The European Organization for Nuclear Research, 1211 Meyrin, Switzerland}
\newcommand{\CIEMAT}{CIEMAT, Centro de Investigaciones Energ{\'e}ticas, Medioambientales y Tecnol{\'o}gicas, E-28040 Madrid, Spain}
\newcommand{\CUSB}{Central University of South Bihar, Gaya {\textendash} 824236, India }
\newcommand{\CalBerkeley}{University of California Berkeley, Berkeley, CA 94720, USA}
\newcommand{\CalDavis}{University of California Davis, Davis, CA 95616, USA}
\newcommand{\CalIrvine}{University of California Irvine, Irvine, CA 92697, USA}
\newcommand{\CalLosangeles}{University of California Los Angeles, Los Angeles, CA 90095, USA}
\newcommand{\CalRiverside}{University of California Riverside, Riverside CA 92521, USA}
\newcommand{\CalSantabarbara}{University of California Santa Barbara, Santa Barbara, California 93106 USA}
\newcommand{\Caltech}{California Institute of Technology, Pasadena, CA 91125, USA}
\newcommand{\Cambridge}{University of Cambridge, Cambridge CB3 0HE, United Kingdom}
\newcommand{\Campinas}{Universidade Estadual de Campinas, Campinas - SP, 13083-970, Brazil}
\newcommand{\CataniaUniversitadi}{Universit{\`a} di Catania, 2 - 95131 Catania, Italy}
\newcommand{\Charles}{Institute f Particle and Nuclear Physics of the Faculty of Mathematics and Physics of the Charles University, 180 00 Prague 8, Czech Republic}
\newcommand{\Chicago}{University of Chicago, Chicago, IL 60637, USA}
\newcommand{\ChungAng}{Chung-Ang University, Seoul 06974, South Korea}
\newcommand{\Cincinnati}{University of Cincinnati, Cincinnati, OH 45221, USA}
\newcommand{\Cinvestav}{Centro de Investigaci{\'o}n y de Estudios Avanzados del Instituto Polit{\'e}cnico Nacional (Cinvestav), Mexico City, Mexico}
\newcommand{\Colima}{Universidad de Colima, Colima, Mexico}
\newcommand{\ColoradoBoulder}{University of Colorado Boulder, Boulder, CO 80309, USA}
\newcommand{\ColoradoState}{Colorado State University, Fort Collins, CO 80523, USA}
\newcommand{\Columbia}{Columbia University, New York, NY 10027, USA}
\newcommand{\CzechAcademyofSciences}{Institute of Physics, Czech Academy of Sciences, 182 00 Prague 8, Czech Republic}
\newcommand{\CzechTechnical}{Czech Technical University, 115 19 Prague 1, Czech Republic}
\newcommand{\DakotaState}{Dakota State University, Madison, SD 57042, USA}
\newcommand{\Dallas}{University of Dallas, Irving, TX 75062-4736, USA}
\newcommand{\DannecyleVieux}{Labratoire d'Annecy-le-Vieux de Physique des Particules, CNRS/IN2P3 and Universit{\'e} Savoie Mont Blanc, 74941 Annecy-le-Vieux, France}
\newcommand{\Daresbury}{Daresbury Laboratory, Cheshire WA4 4AD, United Kingdom}
\newcommand{\Drexel}{Drexel University, Philadelphia, PA 19104, USA}
\newcommand{\Duke}{Duke University, Durham, NC 27708, USA}
\newcommand{\Durham}{Durham University, Durham DH1 3LE, United Kingdom}
\newcommand{\EIA}{Universidad EIA, Antioquia, Colombia}
\newcommand{\ETH}{ETH Zurich, Zurich, Switzerland}
\newcommand{\Edinburgh}{University of Edinburgh, Edinburgh EH8 9YL, United Kingdom}
\newcommand{\FCULport}{Faculdade de Ci{\^e}ncias da Universidade de Lisboa - FCUL, 1749-016 Lisboa, Portugal}
\newcommand{\FederaldeAlfenas}{Universidade Federal de Alfenas, Po{\c{c}}os de Caldas - MG, 37715-400, Brazil}
\newcommand{\FederaldeGoias}{Universidade Federal de Goias, Goiania, GO 74690-900, Brazil}
\newcommand{\FederaldeSaoCarlos}{Universidade Federal de S{\~a}o Carlos, Araras - SP, 13604-900, Brazil}
\newcommand{\FederaldoABC}{Universidade Federal do ABC, Santo Andr{\'e} - SP, 09210-580 Brazil}
\newcommand{\FederaldoRio}{Universidade Federal do Rio de Janeiro,  Rio de Janeiro - RJ, 21941-901, Brazil}
\newcommand{\Fermi}{Fermi National Accelerator Laboratory, Batavia, IL 60510, USA}
\newcommand{\Florida}{University of Florida, Gainesville, FL 32611-8440, USA}
\newcommand{\Fluminense}{Fluminense Federal University, 9 Icara{\'\i} Niter{\'o}i - RJ, 24220-900, Brazil }
\newcommand{\Genova}{Universit{\`a} degli Studi di Genova, Genova, Italy}
\newcommand{\Georgian}{Georgian Technical University, Tbilisi, Georgia}
\newcommand{\GranSasso}{Gran Sasso Science Institute, L'Aquila, Italy}
\newcommand{\GranSassoLab}{Laboratori Nazionali del Gran Sasso, L'Aquila AQ, Italy}
\newcommand{\Granada}{University of Granada {\&} CAFPE, 18002 Granada, Spain}
\newcommand{\Grenoble}{University Grenoble Alpes, CNRS, Grenoble INP, LPSC-IN2P3, 38000 Grenoble, France}
\newcommand{\Guanajuato}{Universidad de Guanajuato, Guanajuato, C.P. 37000, Mexico}
\newcommand{\Harish}{Harish-Chandra Research Institute, Jhunsi, Allahabad 211 019, India}
\newcommand{\Harvard}{Harvard University, Cambridge, MA 02138, USA}
\newcommand{\Hawaii}{University of Hawaii, Honolulu, HI 96822, USA}
\newcommand{\Houston}{University of Houston, Houston, TX 77204, USA}
\newcommand{\Hyderabad}{University of  Hyderabad, Gachibowli, Hyderabad - 500 046, India}
\newcommand{\IFAE}{Institut de F{\`\i}sica d'Altes Energies, Barcelona, Spain}
\newcommand{\IFIC}{Instituto de Fisica Corpuscular, 46980 Paterna, Valencia, Spain}
\newcommand{\INFNBologna}{Istituto Nazionale di Fisica Nucleare Sezione di Bologna, 40127 Bologna BO, Italy}
\newcommand{\INFNCatania}{Istituto Nazionale di Fisica Nucleare Sezione di Catania, I-95123 Catania, Italy}
\newcommand{\INFNGenova}{Istituto Nazionale di Fisica Nucleare Sezione di Genova, 16146 Genova GE, Italy}
\newcommand{\INFNLecce}{Istituto Nazionale di Fisica Nucleare Sezione di Lecce, 73100 - Lecce, Italy}
\newcommand{\INFNMilanBicocca}{Istituto Nazionale di Fisica Nucleare Sezione di Milano Bicocca, 3 - I-20126 Milano, Italy}
\newcommand{\INFNMilano}{Istituto Nazionale di Fisica Nucleare Sezione di Milano, 20133 Milano, Italy}
\newcommand{\INFNNapoli}{Istituto Nazionale di Fisica Nucleare Sezione di Napoli, I-80126 Napoli, Italy}
\newcommand{\INFNPadova}{Istituto Nazionale di Fisica Nucleare Sezione di Padova, 35131 Padova, Italy}
\newcommand{\INFNPavia}{Istituto Nazionale di Fisica Nucleare Sezione di Pavia,  I-27100 Pavia, Italy}
\newcommand{\INFNSud}{Istituto Nazionale di Fisica Nucleare Laboratori Nazionali del Sud, 95123 Catania, Italy}
\newcommand{\INR}{Institute for Nuclear Research of the Russian Academy of Sciences, Moscow 117312, Russia}
\newcommand{\IPLyon}{Institut de Physique des 2 Infinis de Lyon, 69622 Villeurbanne, France}
\newcommand{\IPM}{Institute for Research in Fundamental Sciences, Tehran, Iran}
\newcommand{\ISTlisboa}{Instituto Superior T{\'e}cnico - IST, Universidade de Lisboa, Portugal}
\newcommand{\Idaho}{Idaho State University, Pocatello, ID 83209, USA}
\newcommand{\Illinoisinstitute}{Illinois Institute of Technology, Chicago, IL 60616, USA}
\newcommand{\Imperial}{Imperial College of Science Technology and Medicine, London SW7 2BZ, United Kingdom}
\newcommand{\IndGuwahati}{Indian Institute of Technology Guwahati, Guwahati, 781 039, India}
\newcommand{\IndHyderabad}{Indian Institute of Technology Hyderabad, Hyderabad, 502285, India}
\newcommand{\Indiana}{Indiana University, Bloomington, IN 47405, USA}
\newcommand{\Ingenieria}{Universidad Nacional de Ingenier{\'\i}a, Lima 25, Per{\'u}}
\newcommand{\Iowa}{University of Iowa, Iowa City, IA 52242, USA}
\newcommand{\IowaState}{Iowa State University, Ames, Iowa 50011, USA}
\newcommand{\Iwate}{Iwate University, Morioka, Iwate 020-8551, Japan}
\newcommand{\Jammu}{University of Jammu, Jammu-180006, India}
\newcommand{\Jawaharlal}{Jawaharlal Nehru University, New Delhi 110067, India}
\newcommand{\Jeonbuk}{Jeonbuk National University, Jeonrabuk-do 54896, South Korea}
\newcommand{\Jyvaskyla}{University of Jyvaskyla, FI-40014, Finland}
\newcommand{\KEK}{High Energy Accelerator Research Organization (KEK), Ibaraki, 305-0801, Japan}
\newcommand{\KISTI}{Korea Institute of Science and Technology Information, Daejeon, 34141, South Korea}
\newcommand{\KL}{K L University, Vaddeswaram, Andhra Pradesh 522502, India}
\newcommand{\Kansasstate}{Kansas State University, Manhattan, KS 66506, USA}
\newcommand{\Kavli}{Kavli Institute for the Physics and Mathematics of the Universe, Kashiwa, Chiba 277-8583, Japan}
\newcommand{\Kure}{National Institute of Technology, Kure College, Hiroshima, 737-8506, Japan}
\newcommand{\Kyiv}{Kyiv National University, 01601 Kyiv, Ukraine}
\newcommand{\LIP}{Laborat{\'o}rio de Instrumenta{\c{c}}{\~a}o e F{\'\i}sica Experimental de Part{\'\i}culas, 1649-003 Lisboa and 3004-516 Coimbra, Portugal}
\newcommand{\Lal}{Laboratoire de l'Acc{\'e}l{\'e}rateur Lin{\'e}aire, 91440 Orsay, France}
\newcommand{\Lancaster}{ Lancaster University, Lancaster LA1 4YB, United Kingdom}
\newcommand{\LawrenceBerkeley}{Lawrence Berkeley National Laboratory, Berkeley, CA 94720, USA}
\newcommand{\Liverpool}{University of Liverpool, L69 7ZE, Liverpool, United Kingdom}
\newcommand{\LosAlmos}{Los Alamos National Laboratory, Los Alamos, NM 87545, USA}
\newcommand{\Louisanastate}{Louisiana State University, Baton Rouge, LA 70803, USA}
\newcommand{\Lucknow}{University of Lucknow, Uttar Pradesh 226007, India}
\newcommand{\Madrid}{Madrid Autonoma University and IFT UAM/CSIC, 28049 Madrid, Spain}
\newcommand{\Manchester}{University of Manchester, Manchester M13 9PL, United Kingdom}
\newcommand{\Massinsttech}{Massachusetts Institute of Technology, Cambridge, MA 02139, USA}
\newcommand{\Michigan}{University of Michigan, Ann Arbor, MI 48109, USA}
\newcommand{\Michiganstate}{Michigan State University, East Lansing, MI 48824, USA}
\newcommand{\MilanoBicocca}{Universit{\`a} del Milano-Bicocca, 20126 Milano, Italy}
\newcommand{\MilanoUniv}{Universit{\`a} degli Studi di Milano, I-20133 Milano, Italy}
\newcommand{\Minnduluth}{University of Minnesota Duluth, Duluth, MN 55812, USA}
\newcommand{\Minntwin}{University of Minnesota Twin Cities, Minneapolis, MN 55455, USA}
\newcommand{\Mississippi}{University of Mississippi, University, MS 38677 USA}
\newcommand{\Newmexico}{University of New Mexico, Albuquerque, NM 87131, USA}
\newcommand{\Niewodniczanski}{H. Niewodnicza{\'n}ski Institute of Nuclear Physics, Polish Academy of Sciences, Cracow, Poland}
\newcommand{\Nikhef}{Nikhef National Institute of Subatomic Physics, 1098 XG Amsterdam, Netherlands}
\newcommand{\Northdakota}{University of North Dakota, Grand Forks, ND 58202-8357, USA}
\newcommand{\Northernillinois}{Northern Illinois University, DeKalb, Illinois 60115, USA}
\newcommand{\Northwestern}{Northwestern University, Evanston, Il 60208, USA}
\newcommand{\NotreDame}{University of Notre Dame, Notre Dame, IN 46556, USA}
\newcommand{\Ohiostate}{Ohio State University, Columbus, OH 43210, USA}
\newcommand{\OregonState}{Oregon State University, Corvallis, OR 97331, USA}
\newcommand{\Oxford}{University of Oxford, Oxford, OX1 3RH, United Kingdom}
\newcommand{\PacificNorthwest}{Pacific Northwest National Laboratory, Richland, WA 99352, USA}
\newcommand{\Padova}{Universt{\`a} degli Studi di Padova, I-35131 Padova, Italy}
\newcommand{\Parisuniversite}{Universit{\'e} de Paris, CNRS, Astroparticule et Cosmologie, F-75006, Paris, France}
\newcommand{\Pavia}{Universit{\`a} degli Studi di Pavia, 27100 Pavia PV, Italy}
\newcommand{\Penn}{University of Pennsylvania, Philadelphia, PA 19104, USA}
\newcommand{\PennState}{Pennsylvania State University, University Park, PA 16802, USA}
\newcommand{\PhysicalResearchLaboratory}{Physical Research Laboratory, Ahmedabad 380 009, India}
\newcommand{\Pisa}{Universit{\`a} di Pisa, I-56127 Pisa, Italy}
\newcommand{\Pitt}{University of Pittsburgh, Pittsburgh, PA 15260, USA}
\newcommand{\Pontificia}{Pontificia Universidad Cat{\'o}lica del Per{\'u}, Lima, Per{\'u}}
\newcommand{\PuertoRico}{University of Puerto Rico, Mayaguez 00681, Puerto Rico, USA}
\newcommand{\Punjab}{Punjab Agricultural University, Ludhiana 141004, India}
\newcommand{\Radboud}{Radboud University, NL-6525 AJ Nijmegen, Netherlands}
\newcommand{\Rochester}{University of Rochester, Rochester, NY 14627, USA}
\newcommand{\Royalholloway}{Royal Holloway College London, TW20 0EX, United Kingdom}
\newcommand{\Rutgers}{Rutgers University, Piscataway, NJ, 08854, USA}
\newcommand{\Rutherford}{STFC Rutherford Appleton Laboratory, Didcot OX11 0QX, United Kingdom}
\newcommand{\SLAC}{SLAC National Accelerator Laboratory, Menlo Park, CA 94025, USA}
\newcommand{\SURF}{Sanford Underground Research Facility, Lead, SD, 57754, USA}
\newcommand{\Salento}{Universit{\`a} del Salento, 73100 Lecce, Italy}
\newcommand{\SergioArboleda}{Universidad Sergio Arboleda, 11022 Bogot{\'a}, Colombia}
\newcommand{\Sheffield}{University of Sheffield, Sheffield S3 7RH, United Kingdom}
\newcommand{\SouthDakotaSchool}{South Dakota School of Mines and Technology, Rapid City, SD 57701, USA}
\newcommand{\SouthDakotaState}{South Dakota State University, Brookings, SD 57007, USA}
\newcommand{\Southcarolina}{University of South Carolina, Columbia, SC 29208, USA}
\newcommand{\SouthernMethodist}{Southern Methodist University, Dallas, TX 75275, USA}
\newcommand{\StonyBrook}{Stony Brook University, SUNY, Stony Brook, New York 11794, USA}
\newcommand{\Sussex}{University of Sussex, Brighton, BN1 9RH, United Kingdom}
\newcommand{\Syracuse}{Syracuse University, Syracuse, NY 13244, USA}
\newcommand{\Tennknox}{University of Tennessee at Knoxville, TN, 37996, USA}
\newcommand{\TexasAM}{Texas A{\&}M University - Corpus Christi, Corpus Christi, TX 78412, USA}
\newcommand{\TexasArlington}{University of Texas at Arlington, Arlington, TX 76019, USA}
\newcommand{\Texasaustin}{University of Texas at Austin, Austin, TX 78712, USA}
\newcommand{\Toronto}{University of Toronto, Toronto, Ontario M5S 1A1, Canada}
\newcommand{\Tufts}{Tufts University, Medford, MA 02155, USA}
\newcommand{\Unifesp}{Universidade Federal de S{\~a}o Paulo, 09913-030, S{\~a}o Paulo, Brazil}
\newcommand{\UniversityCollegeLondon}{University College London, London, WC1E 6BT, United Kingdom}
\newcommand{\ValleyCity}{Valley City State University, Valley City, ND 58072, USA}
\newcommand{\VariableEnergy}{Variable Energy Cyclotron Centre, 700 064 West Bengal, India}
\newcommand{\VirginiaTech}{Virginia Tech, Blacksburg, VA 24060, USA}
\newcommand{\Warsaw}{University of Warsaw, 00-927 Warsaw, Poland}
\newcommand{\Warwick}{University of Warwick, Coventry CV4 7AL, United Kingdom}
\newcommand{\Wichita}{Wichita State University, Wichita, KS 67260, USA}
\newcommand{\WilliamMary}{William and Mary, Williamsburg, VA 23187, USA}
\newcommand{\Wisconsin}{University of Wisconsin Madison, Madison, WI 53706, USA}
\newcommand{\Yale}{Yale University, New Haven, CT 06520, USA}
\newcommand{\Yerevan}{Yerevan Institute for Theoretical Physics and Modeling, Yerevan 0036, Armenia}
\newcommand{\York}{York University, Toronto M3J 1P3, Canada}

%Instutions in alphabetical order
\affiliation{\Amsterdam}
\affiliation{\Antananarivo}
\affiliation{\AntonioNarino}
\affiliation{\Argonne}
\affiliation{\Arizona}
\affiliation{\Asuncion}
\affiliation{\Athens}
\affiliation{\Atlantico}
\affiliation{\Banaras}
\affiliation{\Basel}
\affiliation{\Bern}
\affiliation{\Beykent}
\affiliation{\Birmingham}
\affiliation{\BolognaUniversity}
\affiliation{\Boston}
\affiliation{\Bristol}
\affiliation{\Brookhaven}
\affiliation{\Bucharest}
\affiliation{\CBPF}
\affiliation{\CEASaclay}
\affiliation{\CERN}
\affiliation{\CIEMAT}
\affiliation{\CUSB}
\affiliation{\CalBerkeley}
\affiliation{\CalDavis}
\affiliation{\CalIrvine}
\affiliation{\CalLosangeles}
\affiliation{\CalRiverside}
\affiliation{\CalSantabarbara}
\affiliation{\Caltech}
\affiliation{\Cambridge}
\affiliation{\Campinas}
\affiliation{\CataniaUniversitadi}
\affiliation{\Charles}
\affiliation{\Chicago}
\affiliation{\ChungAng}
\affiliation{\Cincinnati}
\affiliation{\Cinvestav}
\affiliation{\Colima}
\affiliation{\ColoradoBoulder}
\affiliation{\ColoradoState}
\affiliation{\Columbia}
\affiliation{\CzechAcademyofSciences}
\affiliation{\CzechTechnical}
\affiliation{\DakotaState}
\affiliation{\Dallas}
\affiliation{\DannecyleVieux}
\affiliation{\Daresbury}
\affiliation{\Drexel}
\affiliation{\Duke}
\affiliation{\Durham}
\affiliation{\EIA}
\affiliation{\ETH}
\affiliation{\Edinburgh}
\affiliation{\FCULport}
\affiliation{\FederaldeAlfenas}
\affiliation{\FederaldeGoias}
\affiliation{\FederaldeSaoCarlos}
\affiliation{\FederaldoABC}
\affiliation{\FederaldoRio}
\affiliation{\Fermi}
\affiliation{\Florida}
\affiliation{\Fluminense}
\affiliation{\Genova}
\affiliation{\Georgian}
\affiliation{\GranSasso}
\affiliation{\GranSassoLab}
\affiliation{\Granada}
\affiliation{\Grenoble}
\affiliation{\Guanajuato}
\affiliation{\Harish}
\affiliation{\Harvard}
\affiliation{\Hawaii}
\affiliation{\Houston}
\affiliation{\Hyderabad}
\affiliation{\IFAE}
\affiliation{\IFIC}
\affiliation{\INFNBologna}
\affiliation{\INFNCatania}
\affiliation{\INFNGenova}
\affiliation{\INFNLecce}
\affiliation{\INFNMilanBicocca}
\affiliation{\INFNMilano}
\affiliation{\INFNNapoli}
\affiliation{\INFNPadova}
\affiliation{\INFNPavia}
\affiliation{\INFNSud}
\affiliation{\INR}
\affiliation{\IPLyon}
\affiliation{\IPM}
\affiliation{\ISTlisboa}
\affiliation{\Idaho}
\affiliation{\Illinoisinstitute}
\affiliation{\Imperial}
\affiliation{\IndGuwahati}
\affiliation{\IndHyderabad}
\affiliation{\Indiana}
\affiliation{\Ingenieria}
\affiliation{\Iowa}
\affiliation{\IowaState}
\affiliation{\Iwate}
\affiliation{\Jammu}
\affiliation{\Jawaharlal}
\affiliation{\Jeonbuk}
\affiliation{\Jyvaskyla}
\affiliation{\KEK}
\affiliation{\KISTI}
\affiliation{\KL}
\affiliation{\Kansasstate}
\affiliation{\Kavli}
\affiliation{\Kure}
\affiliation{\Kyiv}
\affiliation{\LIP}
\affiliation{\Lal}
\affiliation{\Lancaster}
\affiliation{\LawrenceBerkeley}
\affiliation{\Liverpool}
\affiliation{\LosAlmos}
\affiliation{\Louisanastate}
\affiliation{\Lucknow}
\affiliation{\Madrid}
\affiliation{\Manchester}
\affiliation{\Massinsttech}
\affiliation{\Michigan}
\affiliation{\Michiganstate}
\affiliation{\MilanoBicocca}
\affiliation{\MilanoUniv}
\affiliation{\Minnduluth}
\affiliation{\Minntwin}
\affiliation{\Mississippi}
\affiliation{\Newmexico}
\affiliation{\Niewodniczanski}
\affiliation{\Nikhef}
\affiliation{\Northdakota}
\affiliation{\Northernillinois}
\affiliation{\Northwestern}
\affiliation{\NotreDame}
\affiliation{\Ohiostate}
\affiliation{\OregonState}
\affiliation{\Oxford}
\affiliation{\PacificNorthwest}
\affiliation{\Padova}
\affiliation{\Parisuniversite}
\affiliation{\Pavia}
\affiliation{\Penn}
\affiliation{\PennState}
\affiliation{\PhysicalResearchLaboratory}
\affiliation{\Pisa}
\affiliation{\Pitt}
\affiliation{\Pontificia}
\affiliation{\PuertoRico}
\affiliation{\Punjab}
\affiliation{\Radboud}
\affiliation{\Rochester}
\affiliation{\Royalholloway}
\affiliation{\Rutgers}
\affiliation{\Rutherford}
\affiliation{\SLAC}
\affiliation{\SURF}
\affiliation{\Salento}
\affiliation{\SergioArboleda}
\affiliation{\Sheffield}
\affiliation{\SouthDakotaSchool}
\affiliation{\SouthDakotaState}
\affiliation{\Southcarolina}
\affiliation{\SouthernMethodist}
\affiliation{\StonyBrook}
\affiliation{\Sussex}
\affiliation{\Syracuse}
\affiliation{\Tennknox}
\affiliation{\TexasAM}
\affiliation{\TexasArlington}
\affiliation{\Texasaustin}
\affiliation{\Toronto}
\affiliation{\Tufts}
\affiliation{\Unifesp}
\affiliation{\UniversityCollegeLondon}
\affiliation{\ValleyCity}
\affiliation{\VariableEnergy}
\affiliation{\VirginiaTech}
\affiliation{\Warsaw}
\affiliation{\Warwick}
\affiliation{\Wichita}
\affiliation{\WilliamMary}
\affiliation{\Wisconsin}
\affiliation{\Yale}
\affiliation{\Yerevan}
\affiliation{\York}

\author{B.~Abi}\affiliation{\Oxford}
\author{R.~Acciarri}\affiliation{\Fermi}
\author{M.~A.~Acero}\affiliation{\Atlantico}
\author{G.~Adamov}\affiliation{\Georgian}
\author{D.~Adams}\affiliation{\Brookhaven}
\author{M.~Adinolfi}\affiliation{\Bristol}
\author{Z.~Ahmad}\affiliation{\VariableEnergy}
\author{J.~Ahmed}\affiliation{\Warwick}
\author{T.~Alion}\affiliation{\Sussex}
\author{S.~Alonso Monsalve}\affiliation{\CERN}
% ...

\collaboration{The DUNE Collaboration}
\noaffiliation
\date{\today}

\begin{abstract}
  The Deep Underground Neutrino Experiment (DUNE) is a next-generation long-baseline
  neutrino oscillation experiment consisting of a high-power, broadband neutrino beam,
  a highly capable near detector located on site at Fermilab, in Batavia, Illinois,
  and a massive liquid argon
  time projection chamber (LArTPC) far detector located at the 4850L of Sanford
  Underground Research Facility in Lead, South Dakota. The primary scientific goals
  of the experiment are precise measurements of all the parameters governing long-baseline
  neutrino oscillation in a single experiment, sensitivity to observation of neutrinos
  from a core collapse supernova, and sensitivity to physics beyond the Standard Model,
  including baryon number violating processes.
  DUNE has evaluated expected sensitivity to these physics objectives; these results and
  the details of the simulation studies that have been performed to evaluate these sensitivities
  are presented in the DUNE Physics TDR \cite{Abi:2020evt}.
  The long-baseline physics sensitivity calculations presented in the DUNE TDR,
   and in a related physics paper \cite{Abi:2020qib},  
  rely upon
  simulation of the neutrino beam line, simulation of neutrino interactions in the near
  and far detectors, fully automated event reconstruction and neutrino classification,
  and detailed implementation of systematic uncertainties.
  The purpose of this posting is to provide a simplified summary of the
  simulations that went into this analysis to the community, in order to facilitate
  phenomenological studies of long-baseline
  oscillation at DUNE. Simulated neutrino flux files and a GLoBES configuration describing
  the far detector reconstruction and selection performance are included as ancillary files to
  this posting. A simple analysis using these configurations in GLoBES
  produces sensitivity that is similar, but not identical, to the official DUNE sensitivity.
  DUNE welcomes those interested in
  performing phenomenological work as members of the collaboration, but also recognizes
  the benefit of making these configurations readily available to the wider community. 
\end{abstract}

\maketitle

\section{Introduction}
The physics volume of the Technical Design Report (TDR)\cite{Abi:2020evt}  for the Deep Underground Neutrino
Experiment (DUNE) describes the proposed physics program for DUNE and the results of simulation
studies that have been performed to quantify DUNE's sensitivity to its physics objectives.
The primary scientific objectives of DUNE are
to study long-baseline neutrino oscillation to determine the neutrino mass ordering, to determine whether
CP symmetry is violated in the lepton sector, and to precisely measure the parameters governing neutrino
oscillation to test the three-neutrino paradigm. The DUNE physics program also includes precise measurements
of neutrino interactions, sensitivity to supernova burst neutrinos, and searches for a range of physics
beyond the Standard Model, including sensitivity to baryon number violating processes.

The long-baseline physics sensitivity calculations presented in the DUNE TDR
and the related physics paper \cite{Abi:2020qib}
are based upon detailed
simulations of the neutrino beamline and neutrino interactions in the near and far detectors.
To determine the expected physics sensitivities, a full analysis of simulation data is performed, including
automated signal processing, low-level reconstruction, energy reconstruction, and event classification.
This posting provides the results of some of these simulations for use by
anyone in the community interested in studying long-baseline neutrino oscillation in DUNE.
Simulated neutrino flux files and a GLoBES configuration describing
the far detector reconstruction and selection performance are included as ancillary files to
this posting. On arXiv, the ancillary files may be downloaded individually or as
a gzipped tar file.
The DUNE collaboration requests that any results making use of these files
reference this arXiv posting and the paper
describing the TDR long-baseline oscillation analysis\cite{Abi:2020qib}.

A detailed treatment of systematics, based on the expected variations of individual systematic parameters
describing uncertainties in flux, neutrino interactions, and detector effects, is included in full analysis.
However, for the GLoBES files included in this posting, only simple normalization uncertainties are implemented.
It is important to note that the configurations
provided here are not sufficient to fully reproduce the sophisticated analysis presented in the DUNE
TDR. Rather these configurations provide a simplified summary of the simulation, reconstruction, and
event selection aspects of that analysis as a common starting point for phenomenological studies.
The text in this document is not intended to provide thorough documentation of the details of how the
TDR results were produced; that is provided in the TDR text.
Rather we attempt to briefly summarize the analyses that produce these configurations and
provide documentation of how the configurations may be used.

In Section~\ref{sect:flux}, we describe the simulated fluxes for the LBNF beamline design considered by
the TDR, at both the near and far detectors, in both forward horn current (FHC) and reverse horn current (RHC)
modes. These flux files are provided in the ancillary files in a directory called dune\_flux/.
In Section~\ref{sect:xsec}, we describe the neutrino-nucleus interaction model implemented in the TDR
analysis.
In Section~\ref{sect:ana}, we briefly describe simulation, reconstruction, and selection of the expected event
samples in the far detector.
The results of the far detector analysis and a greatly simplified treatment of systematic uncertainties
after constraint by the near detector are provided in the ancillary files in a directory called
dune\_globes/, containing a GLoBES\cite{globes1,globes2} configuration, which is described in
Section~\ref{sect:globes}.

\section{Flux Simulation}
\label{sect:flux}
The neutrino fluxes used in the TDR were produced using G4LBNF, a Geant4\cite{GEANT4:NIM,GEANT4}-based
simulation of the LBNF beamline from primary proton beam to hadron absorber. Specifically, G4LBNF
version v3r5p4 was used, which was built against Geant4 version 4.10.3.p03.
All simulations used the QGSP\_BERT physics list.

\begin{figure}[!htpb]
  \centering
  \includegraphics[width=0.8\textwidth]{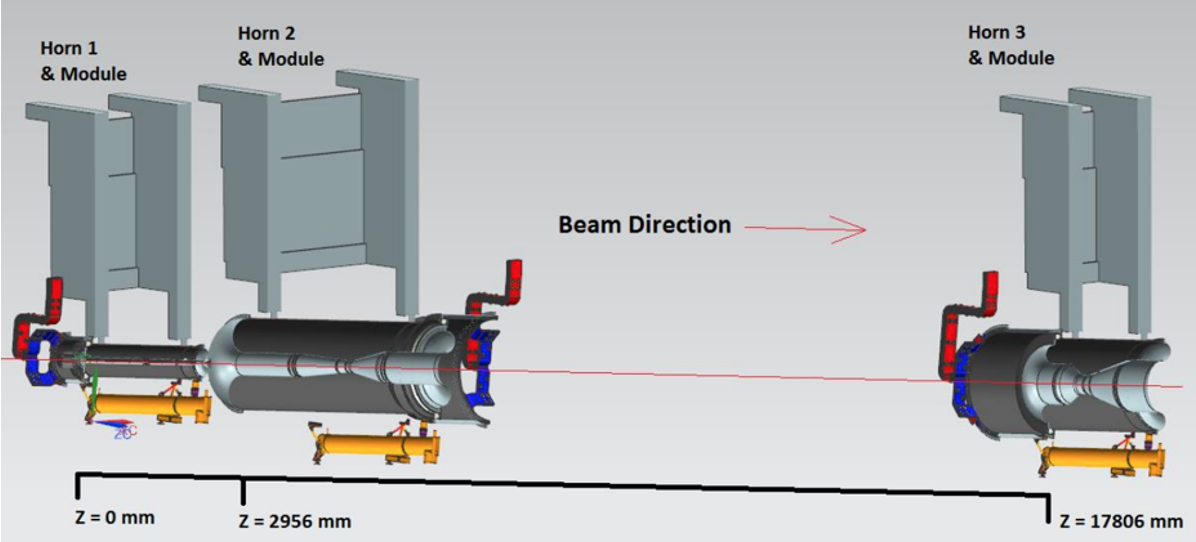}
  \caption{Visualization (C. Crowley, FNAL) of the three-horn focusing system simulated to produce the
    neutrino fluxes used in
  the DUNE TDR and included as an ancillary file with this article.}
  \label{fig:beam_vis}
\end{figure}

G4LBNF is highly configurable to facilitate studies of a variety of beam options.  
The simulation was configured to use a detailed description of the LBNF optimized beam design \cite{beamcdr}.
That design starts with a 1.2-MW, 120-GeV primary proton beam that impinges on a 2.2m long, 16mm diameter
cylindrical graphite target. Hadrons produced in the target are focused by three magnetic horns operated
with 300kA currents. The target chase is followed by a 194m helium-filled decay pipe and a hadron absorber.
The focusing horns can be operated in forward or reverse current configurations, creating neutrino and
antineutrino beams, respectively. As described in \cite{Abi:2020evt}, this design was motivated by a genetic
algorithm used to optimize for CP-violation sensitivity.
The output of the genetic algorithm was a simple design including horn conductor and target shapes,
which was transformed into the detailed conceptual design simulated here by LBNF engineers.
A visualization of the focusing system, as simulated in G4LBNF, is shown in Fig.~\ref{fig:beam_vis}.

The basic output of G4LBNF is a list of all particle decays to neutrinos that occur anywhere along the beamline.
Weights (historically referred to as “importance weights”) are used to reduce the size of the output files by
throwing out a fraction of the relatively common low-energy neutrinos while preserving less numerous
high-energy neutrinos.
To produce neutrino flux distributions at a particular
location, all of the neutrinos in the G4LBNF output file are forced to point toward the specified location and
weighted according to the relative probability that the decay in question would produce a neutrino in
that direction\cite{pavlovic_thesis}.

\begin{figure}[!htpb]
  \centering
  \includegraphics[width=0.4\textwidth]{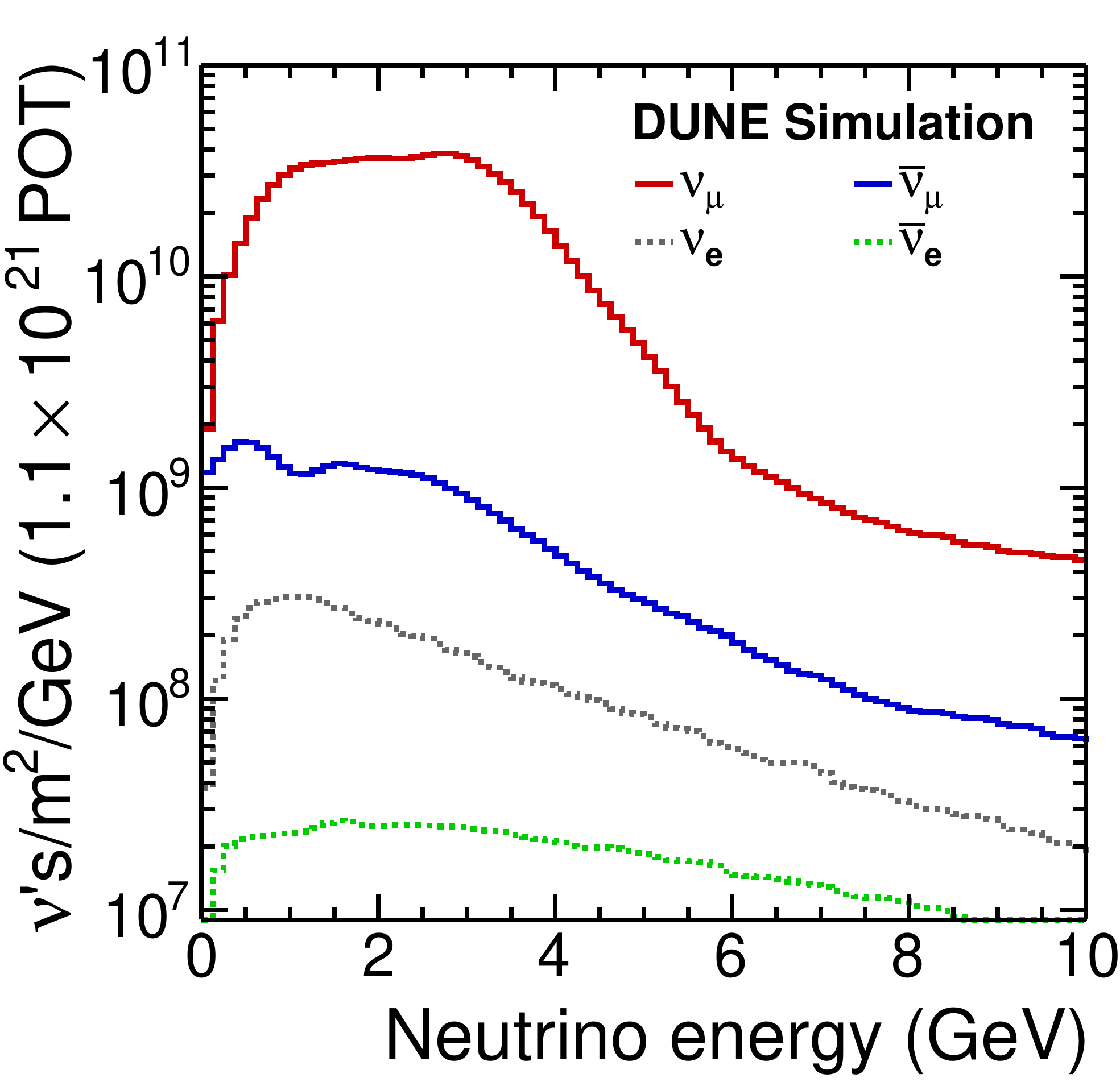}
  \includegraphics[width=0.4\textwidth]{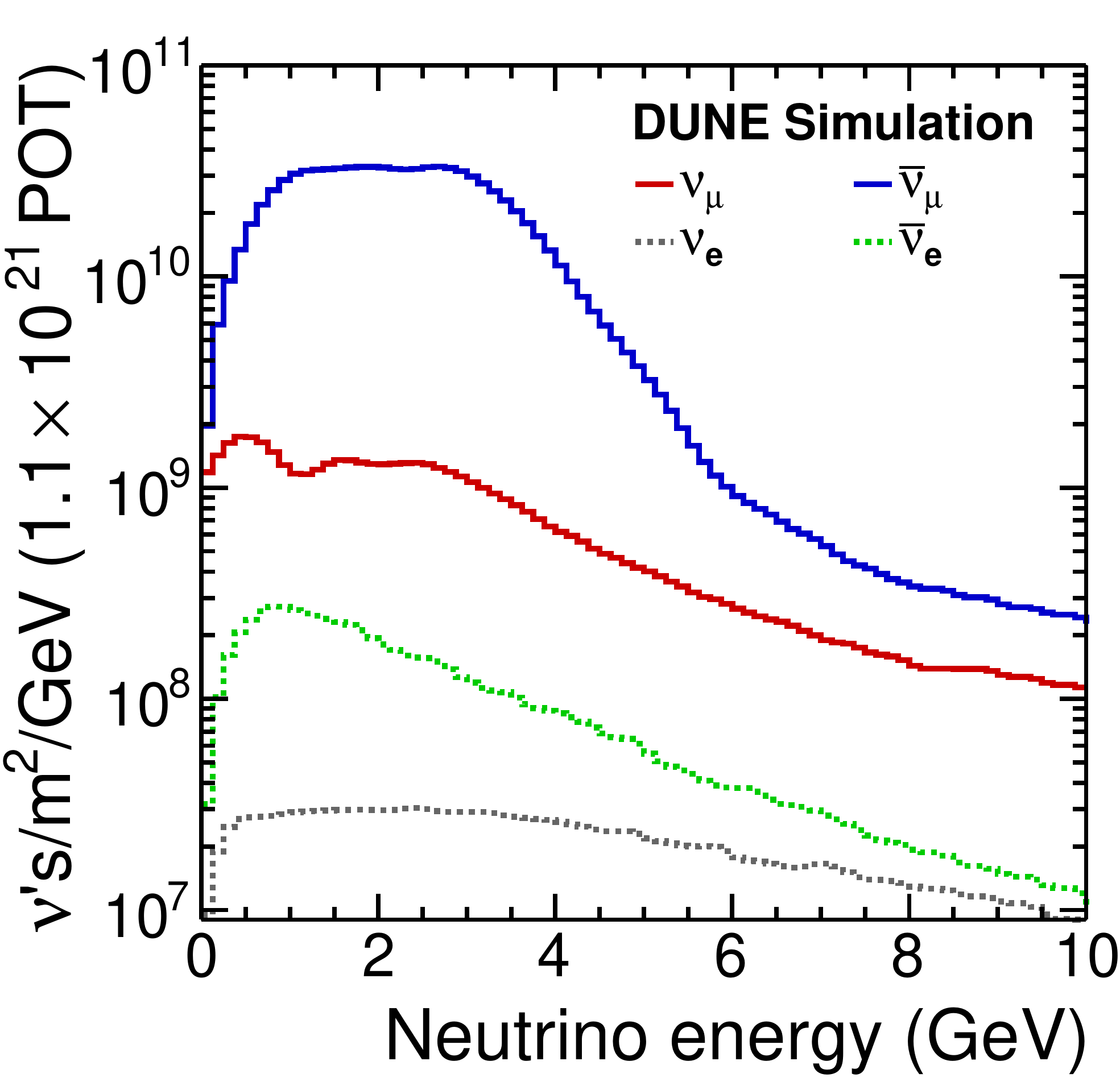}
  \caption{Unoscillated neutrino fluxes at the far detector for neutrino-enhanced, FHC, beam running (left)
    and antineutrino,
    RHC, beam running (right). Figure reproduced from \cite{Abi:2020qib}.}
  \label{fig:beam_flux}
\end{figure}

Fluxes are provided at the center of the near detector (ND), located 574~m downstream
of the start of Horn 1, and
at the far detector (FD), located 1297~km downstream of the start of Horn 1.  Fluxes are available for both
neutrino or ``forward horn current'' mode (FHC) and
antineutrino or ``reverse horn current'' mode (RHC).
The simulated flux distributions at the far detector are shown in Fig.~\ref{fig:beam_flux}.
Each flux is available in two formats: a root file containing flux histograms and a
GLoBES flux input file.  The root files also contain neutral-current and charged-current spectra,
which are obtained by multiplying the flux by GENIE 2.8.4
inclusive cross sections. The flux histograms in the root files have units of neutrinos/$\mathrm{m^2}$/POT.
Note that these histograms have variable bin widths, so discontinuities in the number
of events per bin
are expected.
The GLoBES flux files have units of neutrinos/GeV/$\mathrm{m^2}$/POT. These text files are in the standard
GLoBES format, in which the seven columns correspond to:
$E_{\nu}, \Phi_{\nue}, \Phi_{\numu}, \Phi_{\nutau}, \Phi_{\nuebar}, \Phi_{\numubar},
\mathrm{and} \, \Phi_{\nutaubar}$. The GLoBES far detector flux files are also included as part of the
provided GLoBES configuration.

DUNE plans to measure the flux at various off-axis angles in order to access different neutrino
energy spectra and use this information to reduce the model dependence of the mapping between true
and reconstructed neutrino energy; this concept is referred to as
DUNE PRISM. Flux histograms at off-axis angles have been generated, but are not included in the ancillary
files for this posting. These additional flux files may be found at \cite{lauraspage}.

\section{Cross-Section Simulation}
\label{sect:xsec}

\begin{figure}[!htpb]
  \centering
  \includegraphics[width=0.8\textwidth]{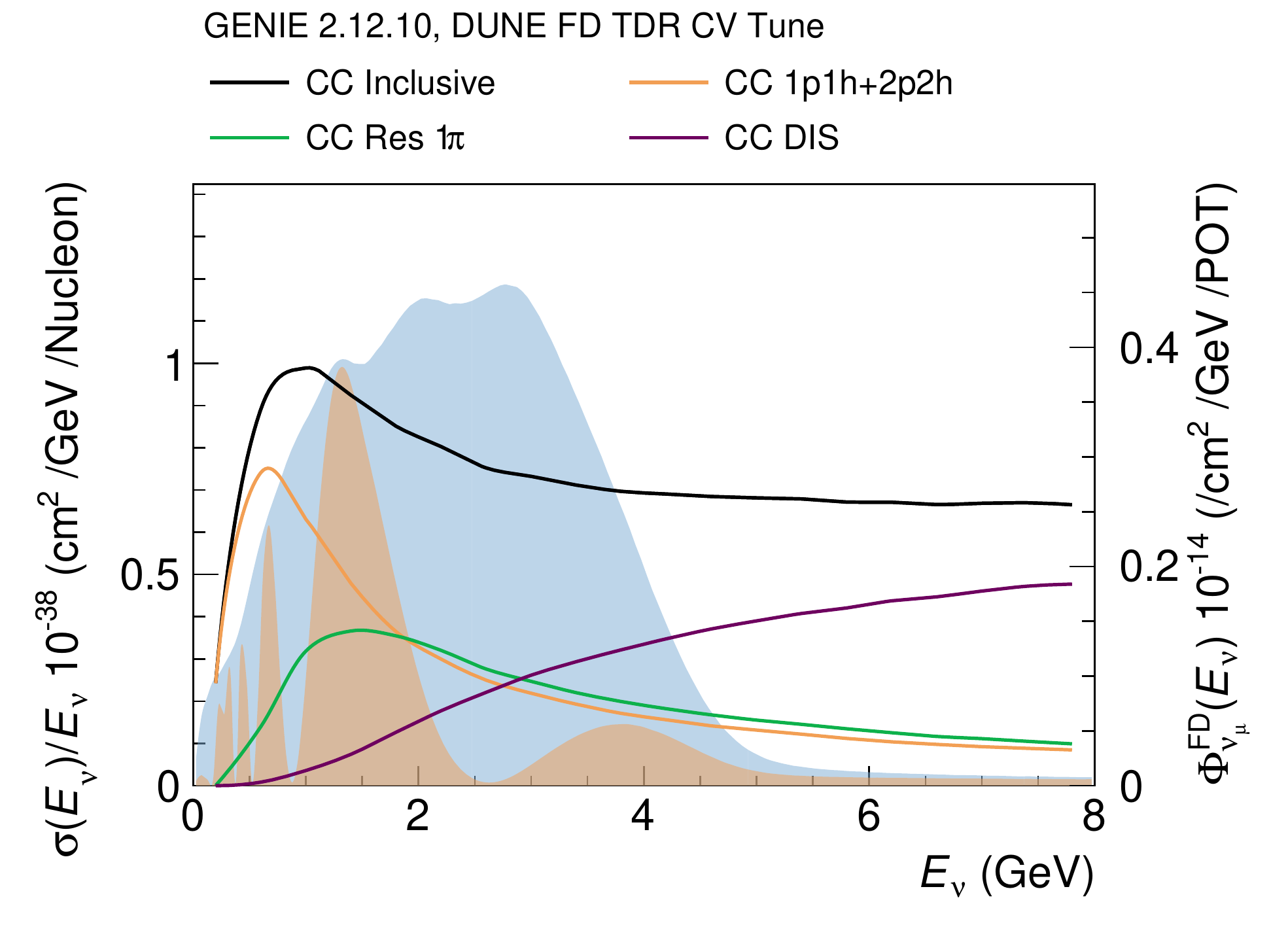}
  \caption{Charged current neutrino interaction cross sections in the DUNE interaction model as
    a function of true neutrino energy. The CC inclusive (black) curve shows the cross-section
    values included in the GLoBES configuration. The blue and orange filled regions show the
    unoscillated and oscillated, respectively, $\numu$ flux at the far detector for reference.}
  \label{fig:xsec}
\end{figure}

A model describing neutrino interactions has been implemented in v2.12.10 of the GENIE
generator \cite{Andreopoulos:2009rq,Andreopoulos:2015wxa}.
Event weights are applied to parameterize cross-section corrections not implemented in this
version of GENIE.
For true charged current quasi-elastic (CCQE) interactions, the shape of the four momentum
transfer is altered according to a
model of the nuclear weak charge screening (RPA) calculated by the Valencia
group \cite{PhysRevC.72.019902}, which results in a strong suppression at low four momentum transfer.
In addition, the rate of resonant and non-resonant single pion production is modified according
to deuterium bubble chamber tunes \cite{Rodrigues:2016xjj}.
The resulting charged-current cross section as a function of energy is shown in Fig.~\ref{fig:xsec}.
For more discussion on the interaction model, the reader is directed to \cite{Abi:2020qib}.
NUISANCE \cite{Stowell:2016jfr} is used to apply weights from the DUNE reweighting framework
to GENIE events and
calculate the total cross sections; this output is used to produce the cross-section text files
included in the GLoBES configuration supplied with this article.
These cross-section text files are in the standard GLoBES format, in which the seven columns correspond to:
$log_{10}E_{\nu}, \sigmahat_{\nue}, \sigmahat_{\numu}, \sigmahat_{\nutau}, \sigmahat_{\nuebar}, \sigmahat_{\numubar}, \mathrm{and} \,
\sigmahat_{\nutaubar}$,
where $\sigmahat(E) = \sigma(E)/E[10^{-38}\frac{\mathrm{cm}^2}{\mathrm{GeV}}]$.
Note that the $\nu_{\tau}$ cross sections are set to zero in these files.

\section{Monte Carlo Analysis}
\label{sect:ana}
As described in \cite{Abi:2020qib}, a full analysis and event selection has been performed on simulated
far detector data.
Simulated data is generated using the flux and cross-section simulation tools described in
the previous section and a Geant4\cite{Agostinelli:2002hh} simulation of the DUNE far detector.

The electronics
response to the ionization
electrons and scintillation light is simulated to produce digitized signals in the wire planes and
photon detectors (PDs) respectively. 
Raw detector signals are processed using algorithms to remove the impact of the LArTPC
electric field and electronics response from the measured signal, to identify hits, and to form clusters of hits
that are matched to form high-level objects such as tracks and showers.

The energy of the incoming neutrino in
charged-current (CC) events is estimated by adding the lepton and hadronic energies reconstructed using the
Pandora toolkit \cite{Marshall:2013bda,Marshall:2012hh}. If the event is selected as $\numu$ CC,
  the muon energy is is estimated from range
of the longest track if it is contained in the detector and from multiple Coulomb scattering if it exits the
detector. Electron and hadron energies are measured calorimetrically, with corrections applied to each hit
charge for recombination and the electron lifetime. An additional correction is made to the hadronic energy to
account for missing energy due to neutral particles and final-state interactions. In the energy range of
$0.5$ to $4$~GeV that is relevant for oscillation measurements, the observed neutrino energy resolution at the far detector
is $\sim$15 to 20\%, depending on lepton flavor and reconstruction method. It is expected that this resolution
could be improved using more sophisticated reconstruction techniques, but those improvements are not
considered in the analysis presented in \cite{Abi:2020qib} or in the GLoBES configurations provided here.

Event classification is carried out through image recognition techniques using a convolutional neural network,
named convolutional visual network (CVN). A detailed description of the CVN architecture is available in
\cite{Abi:2020xvt} and the performance is discussed in \cite{Abi:2020qib}. CVN scores for each interaction
to be a $\numu$ CC or a $\nue$ CC interaction are obtained from a network trained on three million simulated
events. The event selection requirement for an interaction to be included in the $\nue$ CC ($\numu$ CC)
sample is P ($\nue$CC) $>$ 0.85 (P ($\numu$CC) $>$ 0.5), optimized to produce the best sensitivity to CP violation.
Since all of the flavor classification scores must sum to unity, the interactions selected in the two event
selections are completely independent. The same selection criteria are used for both FHC and RHC beam running.
The $\nue$ and $\numu$ selection efficiencies in both FHC and RHC beam modes all exceed 90\% in the neutrino
flux peak.

\section{GLoBES Configuration}
\label{sect:globes}
The GLoBES configuration summarizing the result of the MC-based analysis and facilitating user-generated
sensitivities is provided in the ancillary files in a directory called dune\_globes/.
Cross-section files describing charged-current and neutral-current interactions with argon, generated using
GENIE v2.12.10, with weights applied to match the model choices made in the DUNE simulation, are included in the
configuration.
The true-to-reconstructed smearing matrices and selection efficiency as a function of reconstructed
neutrino energy produced by
the DUNE analysis for various signal
and background modes used by 
GLoBES are included. The selection efficiencies are applied as a ``post-smearing'' efficiency in GLoBES;
i.e., they are applied as a function of reconstructed neutrino energy, such that the configurations
provided reproduce the event rates in Monte Carlo samples generated for
the TDR analysis, including statistical fluctuations.
The naming convention for the channels defined in these files is summarized in
Table~\ref{tab:fastmc_naming_convention}. Note that while smearing and efficiency files are provided in
the configurations for $\nu_\tau$ interactions, the cross-sections for these events are set to zero in the
provided cross-section files, so no $\nu_\tau$ interactions will appear in the event rates when using the
configurations as provided.

\begin{table}[!htb]
  \centering
  \caption{Description of naming convention for channels included in the GLoBES configuration provided in
    ancillary files.
    ``FHC'' and ``RHC'' appear at the beginning of each channel name and
    refer to ``Forward Horn Current'' and ``Reverse Horn Current'' as described in Section~\ref{sect:flux}. Efficiencies are provided
  for both the appearance mode and disappearance mode analyses.}
  \label{tab:fastmc_naming_convention}
  \begin{tabular}{|lcl|} \hline
    Name Includes & Process & Description \\ \hline\hline
    \multicolumn{3}{|l|}{Appearance Mode:} \\
      app\_osc\_nue & $\numu\rightarrow\nue$ (CC) & Electron Neutrino Appearance Signal \\
      app\_osc\_nuebar & $\numubar\rightarrow\nuebar$ (CC) & Electron Antineutrino Appearance Signal\\
      app\_bkg\_nue & $\nue\rightarrow\nue$ (CC) & Intrinsic Beam Electron Neutrino Background\\
      app\_bkg\_nuebar & $\nuebar\rightarrow\nuebar$ (CC) & Intrinsic Beam Electron Antineutrino Background\\
      app\_bkg\_numu & $\numu\rightarrow\numu$ (CC) & Muon Neutrino Charged-Current Background \\
      app\_bkg\_numubar & $\numubar\rightarrow\numubar$ (CC) & Muon Antineutrino Charged-Current Background \\
      app\_bkg\_nutau & $\numu\rightarrow\nutau$ (CC) & Tau Neutrino Appearance Background \\
      app\_bkg\_nutaubar & $\numubar\rightarrow\nutaubar$ (CC) & Tau Antineutrino Appearance Background \\ 
      app\_bkg\_nuNC & $\numu/\nue\rightarrow$~X (NC) & Neutrino Neutral Current Background \\
      app\_bkg\_nubarNC & $\numubar/\nuebar\rightarrow$~X (NC) & Antineutrino Neutral Current Background \\ \hline

      \multicolumn{3}{|l|}{Disappearance Mode:} \\      
      dis\_bkg\_numu & $\numu\rightarrow\numu$ (CC) & Muon Neutrino Charged-Current Signal\\
      dis\_bkg\_numubar & $\numubar\rightarrow\numubar$ (CC) & Muon Antineutrino Charged-Current Signal\\
      dis\_bkg\_nutau & $\numu\rightarrow\nutau$ (CC) & Tau Neutrino Appearance Background \\
      dis\_bkg\_nutaubar & $\numubar\rightarrow\nutaubar$ (CC) & Tau Antineutrino Appearance Background \\
      dis\_bkg\_nuNC & $\numu/\nue\rightarrow$~X (NC) & Neutrino Neutral Current Background \\
      dis\_bkg\_nubarNC & $\numubar/\nuebar\rightarrow$~X (NC) & Antineutrino Neutral Current Background \\ \hline \hline
  \end{tabular}
\end{table}

\begin{figure}[!htpb]
  \centering
  \includegraphics[width=0.45\textwidth]{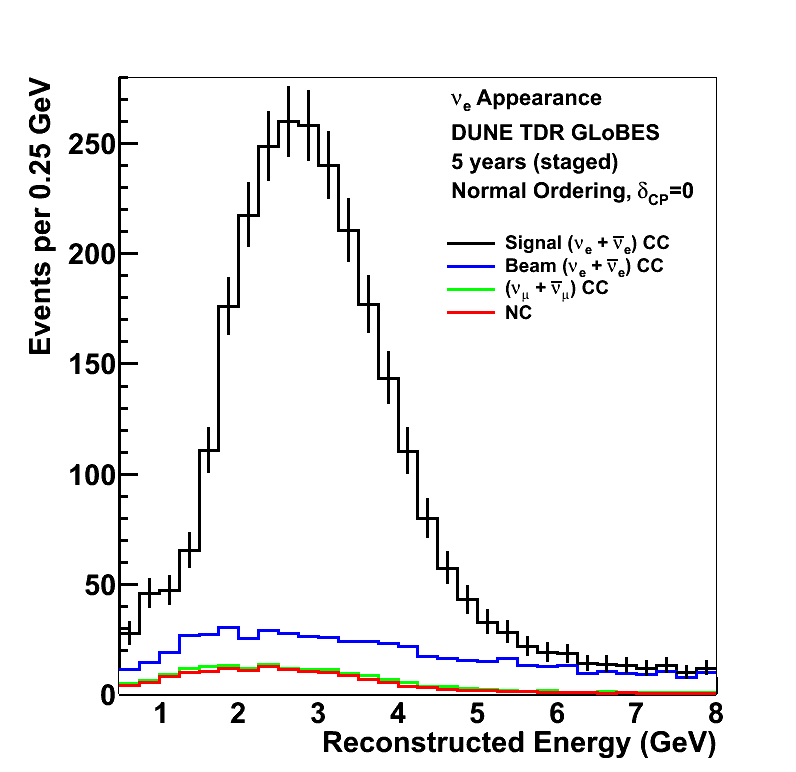}
  \includegraphics[width=0.45\textwidth]{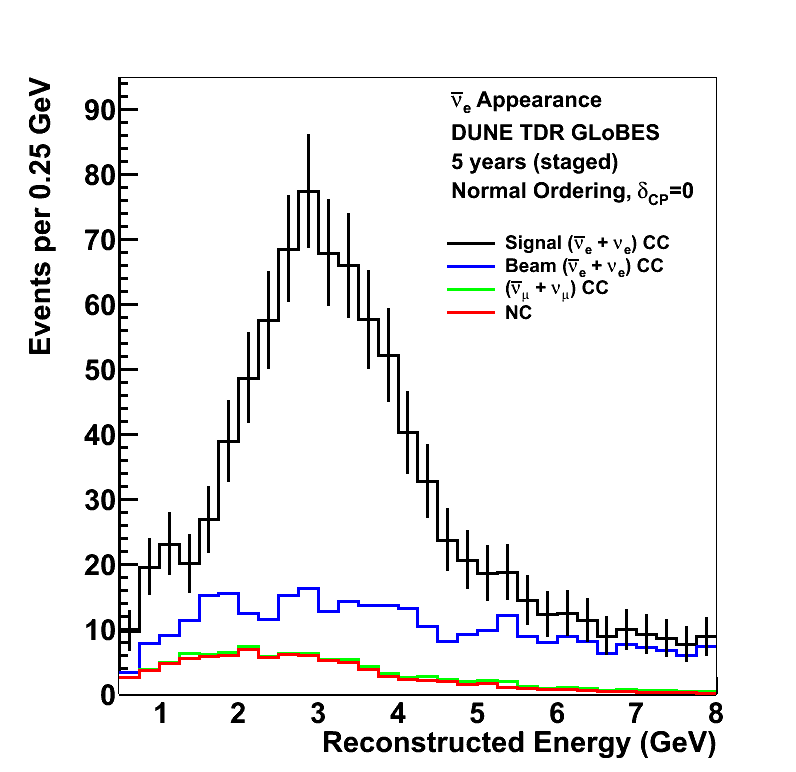}
  \includegraphics[width=0.45\textwidth]{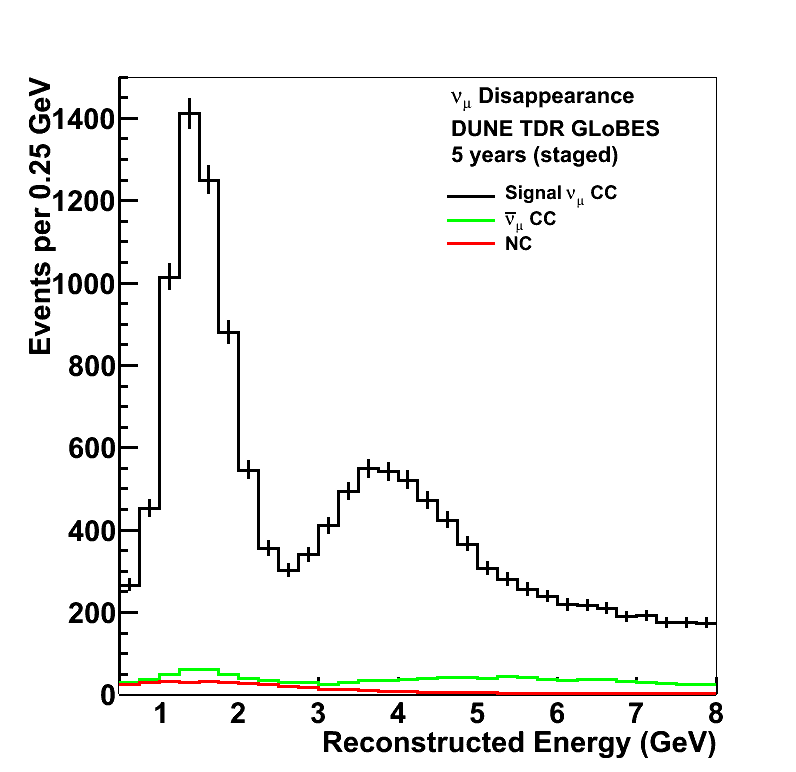}
  \includegraphics[width=0.45\textwidth]{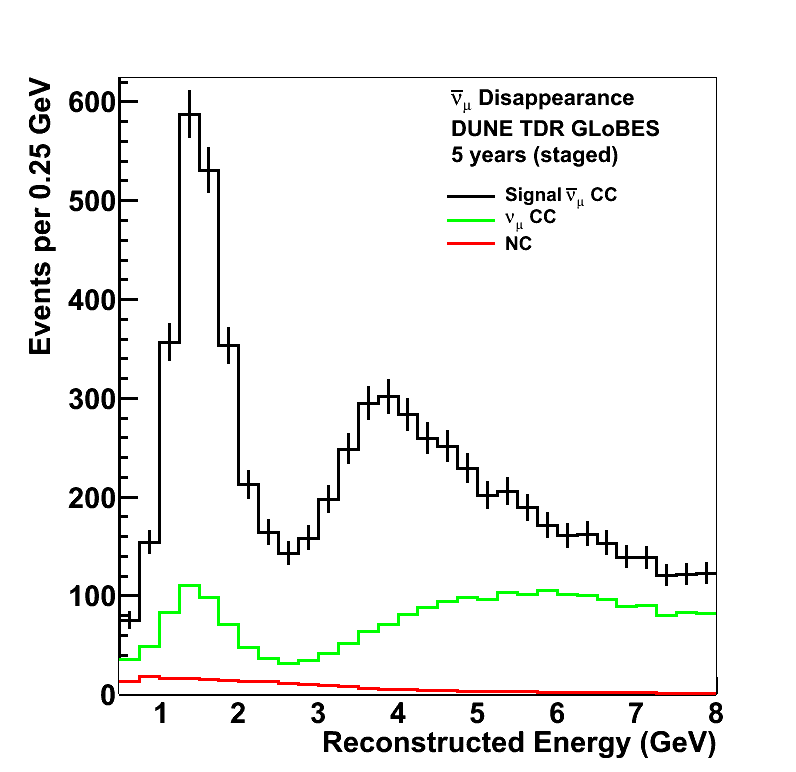}
  \caption{Reconstructed energy distribution of selected $\nue$ CC-like (top) and $\numu$ CC-like (bottom) events,
    assuming 5 years (staged) running in the neutrino-beam mode (left) and antineutrino-beam mode (right), for a
    total of ten years (staged) exposure. True normal ordering is assumed, $\dcp$=0, and all other oscillation
    parameters have the central values given in Table~\ref{tab:oscpar_nufit}.
    Statistical uncertainties are shown on the black histogram.
    Background and signal
    distrubtions are displayed as stacked histograms such that the black histogram represents the full selected
    sample.
    Spectra are
    generated using the GLoBES configuration provided as an ancillary file in this article. }
  \label{fig:spectra}
\end{figure}

\begin{figure}[!htpb]
  \centering
  \includegraphics[width=0.3\textwidth]{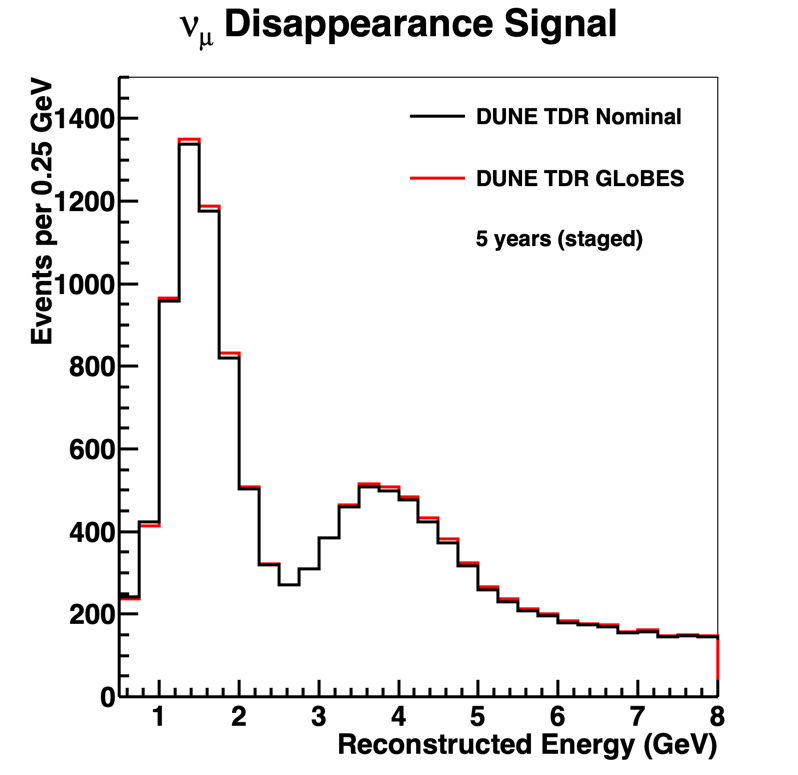}
  \includegraphics[width=0.3\textwidth]{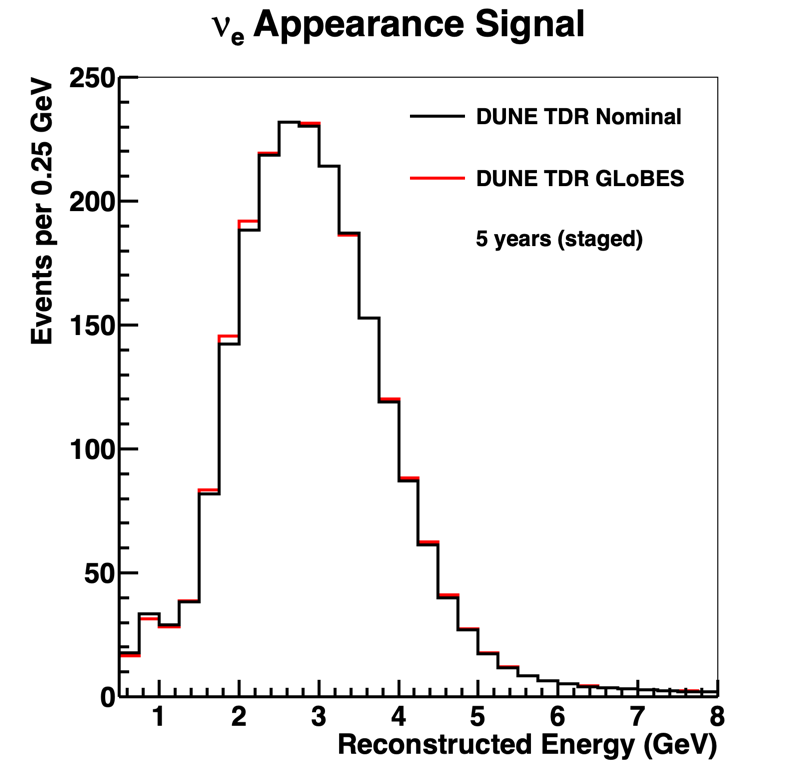}
  \includegraphics[width=0.3\textwidth]{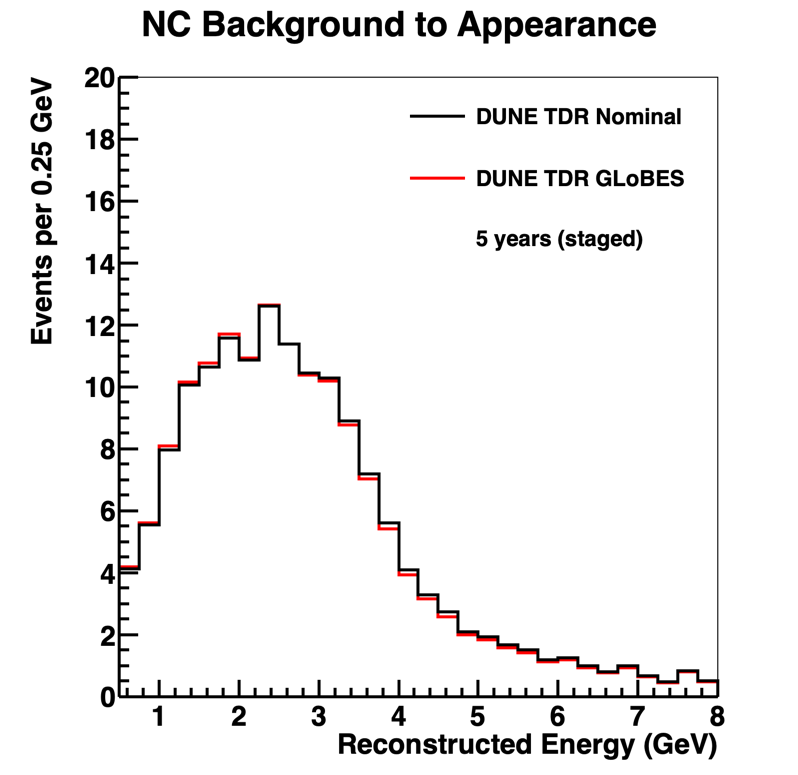}
  \caption{Reconstructed energy distribution of selected true $\numu$ CC (left), $\nue$ CC (middle), and
    neutral current (NC) background to the appearance mode (right) events,
    assuming 5 years (staged) running in the neutrino-beam mode. Event rates from the nominal TDR analysis
    (black histogram) are compared to event rates produced by the GLoBES configuration provided with
    this article (red histogram). True normal ordering is assumed, $\dcp$=0, and all other oscillation
    parameters have the central values given in Table~\ref{tab:oscpar_nufit}.}
  \label{fig:speccomp}
\end{figure}

The GLoBES configuration provided in the ancillary files corresponds to 624 kt-MW-years of exposure: 
6.5 years each of running in neutrino (FHC) and antineutrino (RHC)
mode with a 40-kt fiducial mass far detector, in an 120-GeV, 1.2 MW beam. This is equivalent to ten
years of data collection using the nominal staging assumptions described in \cite{Abi:2020qib}. Conversion between
exposure in kt-MW-years and true years for several nominal exposures is also provided in \cite{Abi:2020qib}.
The sensitivity calculations presented here and in the DUNE nominal analysis \cite{Abi:2020qib}
use oscillation parameters and uncertainties based on the NuFit
4.0\cite{Esteban:2018azc,nufitweb} fit to global neutrino data.
These central values are provided in Table~\ref{tab:oscpar_nufit}.
The matter density is constant and equal to 2.848 $g/cm^{3}$, the average matter density for this baseline
\cite{doi:10.1002/2016JB012887,Roe:2017zdw}.
Figure \ref{fig:spectra} shows the expected DUNE far detector spectra produced by the GLoBES configuration
provided here. These spectra are nearly identical to those produced by the full analysis, as demonstrated in
Fig.~\ref{fig:speccomp}.

\begin{table}[!htb]
\centering
\caption{Central value and relative uncertainty of neutrino oscillation parameters from the
  NuFit 4.0 \cite{Esteban:2018azc,nufitweb}
  global fit to neutrino oscillation data. Because the probability distributions are somewhat non-Gaussian
  (particularly for $\theta_{23}$), the relative uncertainty is computed using 1/6 of the $\pm3\sigma$ allowed range
  from the fit, rather than the 1$\sigma$ range.   For some parameters, the best-fit values
  and uncertainties depend on whether normal mass ordering (NO) or inverted mass ordering (IO) is assumed.}
\label{tab:oscpar_nufit}
\begin{tabular}{|lcc|} \hline \hline
Parameter &    Central Value & Relative Uncertainty \\ \hline
$\theta_{12}$ & 0.5903 & 2.3\% \\ \hline
$\theta_{23}$ (NO) & 0.866  & 4.1\% \\ 
$\theta_{23}$ (IO) & 0.869  & 4.0\% \\ \hline
$\theta_{13}$ (NO) & 0.150  & 1.5\% \\
$\theta_{13}$ (IO) & 0.151  & 1.5\% \\ \hline
$\Delta m^2_{21}$ & 7.39$\times10^{-5}$~eV$^2$ & 2.8\% \\  \hline
$\Delta m^2_{32}$ (NO) & 2.451$\times10^{-3}$~eV$^2$ &  1.3\% \\ 
$\Delta m^2_{31}$ (IO) & -2.512$\times10^{-3}$~eV$^2$ &  1.3\% \\ \hline \hline
\end{tabular}
\end{table}

In all cases, oscillation parameters are allowed to vary in the sensitivity calculations.
The mixing angle $\theta_{13}$ and the solar oscillation parameters, $\theta_{12}$ and $\Delta m^2_{12}$,
are constrained by
Gaussian prior functions with widths defined by the uncertainties in Table~\ref{tab:oscpar_nufit}.
The uncertainty on the matter density is taken to be 2\%.
The GLoBES minimization is performed over both possible values for the $\thetatwothree$ octant and,
in the case of CP violation
sensitivity, both possible values for the neutrino mass ordering.

The $\nue$ and $\nuebar$ signal modes have independent normalization 
uncertainties of 2\% each, while the $\numu$ and $\numubar$ signal modes have independent normalization
uncertainties of 5\%.
The background normalization uncertainties range from 5\% to 20\% and
include correlations among various sources of background; the correlations among the background normalization
parameters can be seen by looking at the @sys\_on\_multiex\_errors\_bg parameters in the GLoBES configurations
provided with this posting.
The choices for signal and background normalization uncertainties
may be customized by changing the parameter values in the file definitions.inc.
The treatment of correlation among uncertainties in this configuration
requires use of GLoBES version 3.2.16, available from the GLoBES website\cite{globesweb}.
Note that while the analysis described in the TDR {\it explicitly} includes selected near detector samples,
the normalization uncertainties here were chosen to {\it implicitly} include the effect of the near
detector and so approximate the expected uncertainty after constraints from the near detector are included.

\begin{figure}[!htpb]
  \centering
  \includegraphics[width=0.4\textwidth]{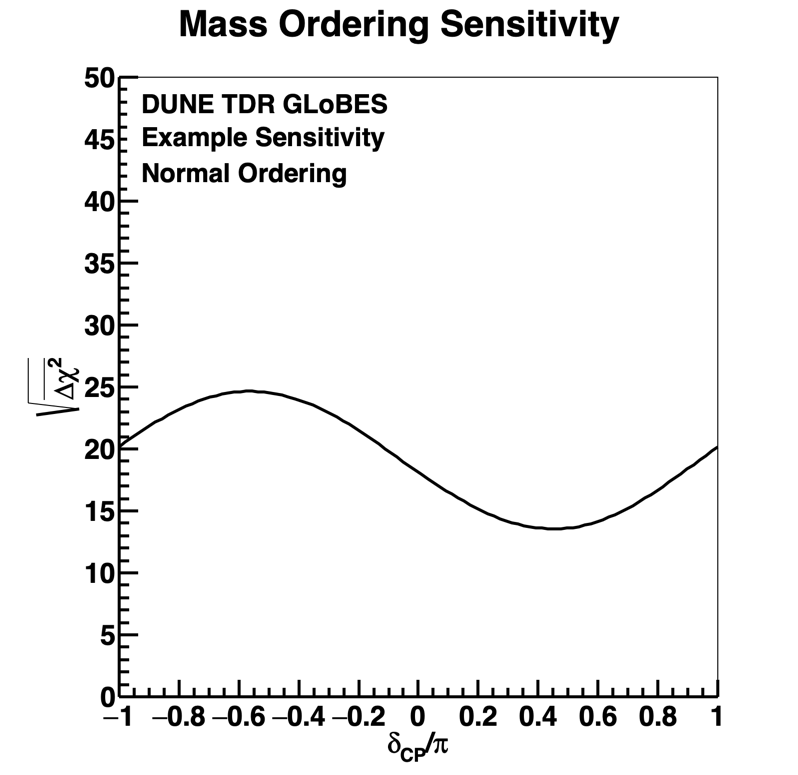}
    \includegraphics[width=0.4\textwidth]{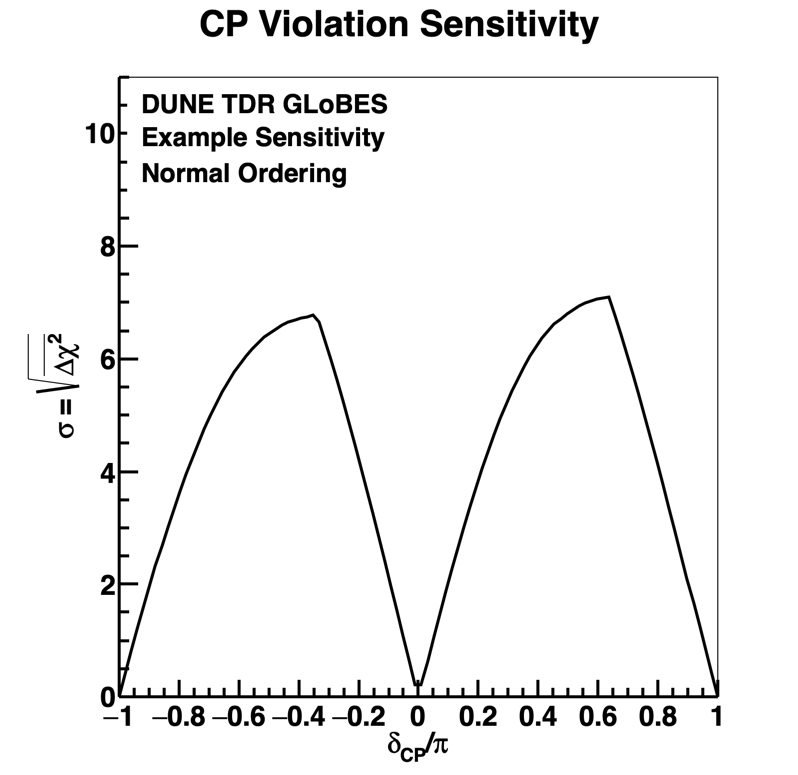}
    \caption{The significance with which the neutrino mass ordering can be determined (left) or CP violation can be
      discovered ($i.e.: \dcp \ne 0$ or $\pi$, right)
      as a function of the value of $\dcp$ for an exposure of 624 kt-MW-years (10 staged years),
      assuming equal exposure in neutrino and antineutrino
      mode and true normal hierarchy, using the provided configuration and parameters given in this document.}
  \label{fig:osc_sens}
\end{figure}

Figure \ref{fig:osc_sens} shows the sensitivity of DUNE's Asimov data to determination
of the neutrino mass ordering and discovery of CP violation, based on the configurations provided here,
assuming an exposure of 624 kt-MW-years. It is important to note that, due to differences in the analysis,
particularly the treatment
of systematic uncertainty, the sensitivity is similar, but not identical, to the official DUNE sensitivity
described in \cite{Abi:2020evt,Abi:2020qib}. The sensitivity curves in Fig.~\ref{fig:osc_sens} are provided only
to assist in validation of user implementation of these configurations.

\section{Summary}
The results of simulations of the LBNF neutrino beamline and a full Monte Carlo simulation and analysis of
expected DUNE far detector neutrino interactions are provided to facilitate phenomenological studies of DUNE
physics sensitivity. The GLoBES configurations provided here produce spectra that are nearly identical to those
used in the nominal DUNE TDR analysis. These configurations produce sensitivity that is similar, but not identical,
to the nominal DUNE TDR analysis for 
neutrino mass ordering and CP violation; the differences are primarily due to a simplified treatment
of systematic uncertainty relative to that in the nominal analysis.
The DUNE collaboration welcomes
those interested in studying DUNE to join the collaboration or to use these configurations independently.
Discussion of any results with the DUNE collaboration, either as a member or a guest, is encouraged. The collaboration
requests that any results making use of the ancillary files reference this arXiv posting and the paper
describing the TDR long-baseline oscillation analysis\cite{Abi:2020qib}.

\bibliographystyle{h-physrev}
\bibliography{tdr_configs_bib}

\end{document}